\documentclass[pra,reprint, twocolumn, subscriptaddress, showpacs]{revtex4}
\usepackage{amsmath}
\usepackage{graphicx}
\usepackage{hyperref}
\usepackage{tabularx}

\begin{document}

    %\preprint{}

    \title{Efficient sympathetic cooling in mixed barium and ytterbium ion chains} 



    \author{Tomasz P. Sakrejda}
    \email[]{Corresponding author: tomaszs@uw.edu}
    %\thanks{}
    \author{Liudmila A. Zhukas}
    \author{Boris B. Blinov}
    \affiliation{Department of Physics, University of Washington, Seattle, Washington 98195, USA}


    \date{\today}

    \begin{abstract}
		
	We study sympathetic cooling of the radial ion motion in a linear RF trap in mixed barium-ytterbium chains. Barium ions are Doppler-cooled, while ytterbium ions are cooled through their interaction with cold barium ions. We estimate the efficiency of sympathetic cooling by measuring the average occupation quantum numbers, and thus the temperature, of all radial normal modes of motion in the ion chain. The full set of orderings in a chain of two barium and two ytterbium ions have been probed, and we show that the average thermal occupation numbers for all chain configurations strongly depend on the trap aspect ratio. We demonstrate efficient sympathetic cooling of all radial normal modes for the trap aspect ratio of approximately 2.9.
    \end{abstract}
    %\pacs{32.60.+i,32.10.Fn,32.30.Bv}
    %\keywords{}
    \maketitle
    \section{Introduction}
    %Begin intro
     \indent Trapped atomic ions are promising candidates technology for a quantum computer with near-perfect qubit initialization and detection efficiency, and high fidelity entangling gates~\cite{HAFFNER2008155}. However, single species trapped ion quantum processors run into several difficulties. First, most high-fidelity Coulomb gates with ions require to keep the ions to be cold~\cite{PhysRevLett.74.4091}. Second, in some architectures, Bell-measurement schemes are used for entanglement between separate ion traps~\cite{nature2007,PhysRevA.89.022317}. These schemes involve repeated excitation of some of the ions in the chain with intense, resonant laser pulses, which leads to decoherence of the neighboring ions. To avoid these complications, mixed-species chains of the trapped ions can be  used~\cite{doi:10.1255/ejms.1408,PhysRevLett.118.053002}. In a mixed-species scheme, one ion type can be used as the \textquotedblleft utility\textquotedblright~species, while the other as the \textquotedblleft logic\textquotedblright~species. The former is used for cooling the entire ion chain and establishing the remote entanglement, and the latter is used for the quantum information storage and manipulation.\linebreak  
    \indent Same-element mixed species schemes using differing isotopes have been proposed \cite{1464-4266-3-1-357,1347-4065-38-4R-2141,2002PhRvA..65d0304B}, but off-resonant scattering of the cooling laser light by the logic species is significant\cite{hucul_modular_2014}\cite{home_memory_2009}, so using different elements is preferable\cite{wang_single-qubit_2017}. However, if species of two different elements are used, the large mass difference can cause the normal mode couplings to become small. This impairs sympathetic cooling and entanglement swapping \cite{PhysRevA.85.043412}. Barium (Ba) and ytterbium (Yb) atomic ions represent a promising choice of the \textquotedblleft utility\textquotedblright/\textquotedblleft logic\textquotedblright~ion pair. Their relative mass difference is fairly small at approximately 25\%. However, as shown in~\cite{PhysRevA.85.043412}, even this small mass difference is sufficient to potentially significantly reduce the radial mode coupling. Thus, a careful experimental study is justified.\linebreak 
    \indent For scaling to larger trapped ion quantum processors, radial modes are preferred to axial modes for local gates~\cite{refId0,PhysRevLett.97.050505}. Using these modes offers several advantages: higher frequencies allow faster gate speeds, performing operations with lasers perpendicular to the ion chain axis makes it easier to address single ions, and gate infidelity due to the ions' thermal motion is reduced by a factor up to  $(\frac{\omega_{r}}{\omega_{a}})^6$, where the $\omega_{r}$ is the radial center-of-mass (COM) mode frequency, and $\omega_{a}$ is the the axial COM mode frequency~\cite{1367-2630-17-10-103025}.\linebreak     
    \indent Unfortunately, in the case of mixed species chains radial modes' coupling is small for typical trapping parameters due to the species' mass difference \cite{PhysRevA.85.043412}. This coupling can be characterized by the largest eigenvector component for each species in a given mode, $\max(\beta_{ij})$, where $\beta_{ij}$ is the eigenvector component of the $j$th ion in the $i$th mode. The coupling is important for cooling, as the rate of sympathetic cooling scales as $\max(\beta_{ij})$ \cite{PhysRevA.61.032310}. \linebreak
    \indent In this paper, we demonstrate efficient sympathetic cooling and good radial vibrational mode coupling in four-ion barium-ytterbium ion chains confined in a linear RF trap. We use the barium ion as a thermometer and find that the coupling for each normal mode depends strongly on the trap aspect ratio. In the strongly coupled case, the measured mode temperatures are consistent with the Doppler cooling limit, while in the weakly coupled case they are one to two orders of magnitude higher.
    \section{Theory}
\begin{figure}[h!!]
\includegraphics[width=3.25in]{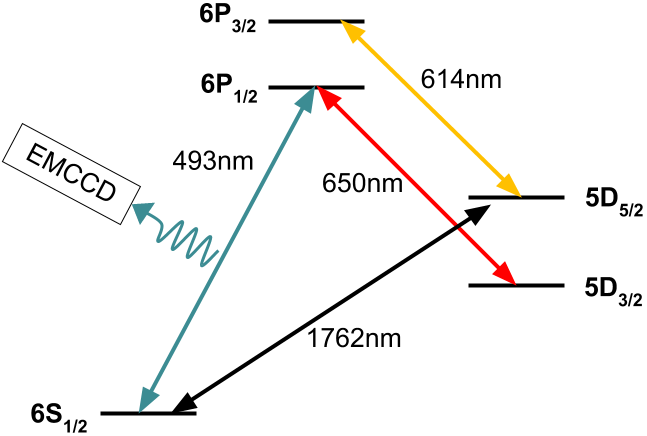}
\caption{(Color online) Ba$^+$ energy level diagram. The 493~nm fluorescence from the ions is imaged onto the EMCCD camera. The ion decays from the $6P_{1/2}$ excited state to the $5D_{3/2}$ metastable state with a branching fraction of $0.25$, and the 650~nm is used to quickly repump from this long-lived state. The 1762~nm laser coherently drives the ion into the $5D_{5/2}$ state where it is \textquotedblleft shelved\textquotedblright~and does not participate in the cooling cycle. The 614~nm laser is used to quickly de-shelve the ion.}
\label{fig:BaLevels}
\end{figure}

    In the mixed-species ion chain only one ion species undergoes direct laser cooling operations. The second species is cooled by the collective motion of the whole chain. Thus, sufficiently strong coupling of the second species ions to all normal modes of motion is necessary for efficient sympathetic cooling. The strength of the ions' coupling to the motional modes is characterized by the normal modes vectors' components, which can be numerically calculated. The cooling rates of each radial mode of our mixed barium-ytterbium ion chain may be calculated analogously to~\cite{PhysRevA.85.043412}. For Doppler cooling, the cooling rate of the $i$th mode is:
    
    \begin{equation}
        \frac{dE_{i}}{dt}\Bigr|_{cool} = \hbar k^2\cos^2{\theta} \frac{I}{I_0}\frac{2\Delta\Gamma}{(1+I/I_0 + (2\Delta/\Gamma)^2)^2}z_i^2\beta_{ij}^2\omega_i^2\mathrm{,}
        \label{eqn:templimit}
    \end{equation}
    
    where k is the wavevector, $\theta$ is the angle between the wavevector and the mode axis, $I$ and $I_0$ are the laser and saturation intensities, $\Delta$ is the detuning, $\Gamma$ is the linewidth, $z_i$ is the normal mode amplitude, and $\omega_i$ is the secular frequency for that mode.\linebreak
    \indent The average occupation number for each radial mode can then be determined from a steady-state equilibrium condition between the heating and cooling mechanisms. When external heating is low and photon recoil dominates, the Doppler limit might be achieved. The estimated heating rate of a single ion $\dot{\bar{n}}$ due to the absorption and emission of the 493~nm photon is of order $10^5$ quanta/second for the secular trap frequency of 1~MHz. The measured external heating rate of a single ion in our trap when cooling lasers are shuttered is of order 2500 quanta/second, which is smaller than diffusion heating.\linebreak
    %%Given that the closest distance from ions to the electrodes in our trap is approximately 250~$\mu$$m$, the estimated heating due to random electric field fluctuations of the patch potential on the trap electrodes at room temperature \cite{PhysRevA.84.023412} is of order 500 quanta/second at 1~MHz.
    \indent In the opposite case when external heating due to random electric field fluctuations of the patch potential on the trap electrodes dominates~\cite{PhysRevA.84.023412}, photon heating can be neglected. Ignoring a summation over the eigenvector components which is of order unity for our chains of up to 4 ions, the temperature limit for a given radial mode can be written in terms of the normal mode occupation number as: 
    \begin{equation}
        \bar{n}_{i} = \max_{\beta_{ij}}\frac{q^2 S_E \Gamma (1 + (2 \Delta / \Gamma )^2 + I/I_0)}{16 \beta_{ij}^2 |\Delta| \hbar^2 \omega_i k^2 \cos^2{\theta}\: I/I_0 } \mathrm{,}
        \label{eqn:templimit}
    \end{equation}
    
    where $S_E$ is the spectral noise density in the trap at the ion site, and $q$ is the charge of a single ion~\cite{PhysRevA.85.043412}. \newline
    \indent In our experiments we extract chains' radial motional occupation numbers by measuring the strength of the $\Delta n = 1$ radial motional sidebands of Ba$^+$ narrow $6S_{1/2}-5D_{5/2}$ \textquotedblleft shelving\textquotedblright~transition near 1762~nm (see Figure~\ref{fig:BaLevels}), relative to the strength of the carrier transition. The Rabi frequency $\Omega_{ij}$ for the $\Delta n=1$ sideband of $i$th normal mode for the $j$th ion in the chain is:
    \begin{equation}
        \Omega_{ij} = \sqrt{n_{i}+1} \beta_{ij} \eta_{i} \Omega_{j} \mathrm{,}
        \label{eqn:sidebandrabi}
    \end{equation}
        
where $n_i$ is the radial motional occupation number for the $i$th mode, $\eta_{j}$ is the Lamb-Dicke parameter, and $\Omega_{j}$ is the carrier Rabi frequency (we allow for spatial dependence of the carrier Rabi frequency due to variation in the laser's beam intesity). Each ordering of ions has a different normal mode decomposition, with both different eigenvectors and different frequencies, so a particular order is maintained for the full duration of an experiment.\linebreak
 \indent To extract the average occupation numbers of different radial modes in the ion chain of $N$ ions, we calculate the transition probability $P_{j}$ for ion $j$ using the Rabi frequency from eqn. \ref{eqn:sidebandrabi}, and sum over over both the full occupation number distribution and over all radial normal modes in the spectrum. Assuming a thermal state, we have:
        %%%%%%%WEAK EXCITATION OF SIDEBANDS%%%%%%%
    \begin{equation}
        \label{eqn:FullWeakExcite}
        P_{j} = \sum_{i=1}^{2N}\sum_{n_i=0}^{\infty}\frac{1}{\bar{n}_i+1}\left(\frac{\bar{n}_i}{\bar{n}_i+1}\right)^{n_{i}}\sin^2(\Omega_{ij}t/2)\mathrm{,}
    \end{equation}

    where $\bar{n}_i$ is the given mode's average occupation number and $t$ is the 1762~nm laser exposure time.
    
\begin{figure}[h!!]
        \includegraphics[width=2.6in]{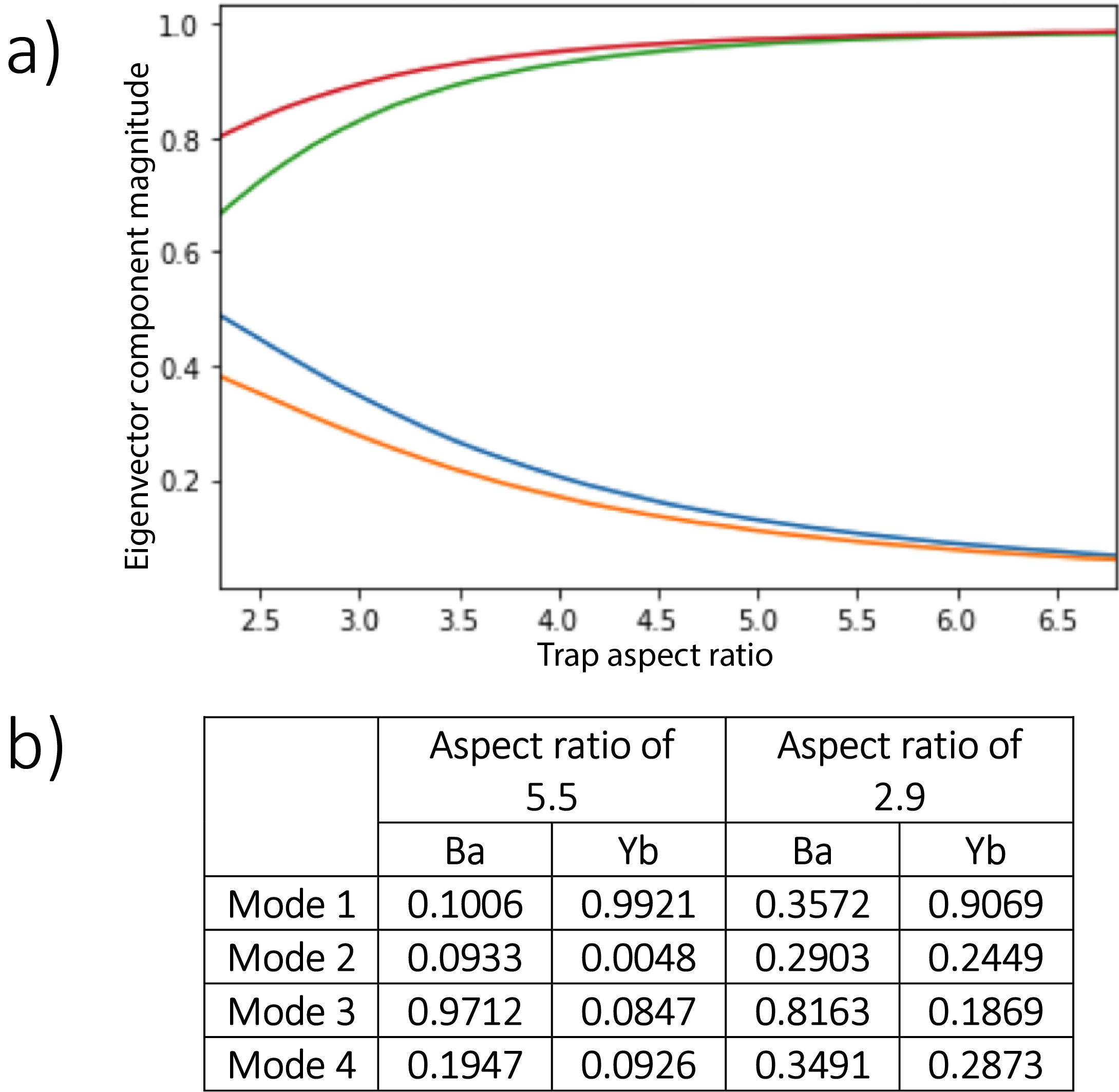}
        \caption{(a) Eigenvector components of the second Ba$^+$ ion in the BYBY chain in all four radial modes along the $x$ principal axis of the trap. The left cutoff is the edge of the zig-zag transition. The two normal modes with larger eigenvectors (two upper lines) couple more strongly to the motion of the lighter Ba ions, while the two normal modes with smaller eigenvectors (two lower lines) couple weakly. As the trap aspect ratio decreases, eigenvector magnitudes with values lower than 0.5 increase and those with values higher than 0.5 decrease. (b) Eigenvector magnitudes of the first ytterbium and the second barium ions for all four radial modes in $x$-direction at the aspect ratios of 5.5 and 2.9. All four ions are in the same BYBY configuration.}
        \label{fig:normal_modes}
    \end{figure}

    To estimate normal mode occupation numbers, we calculate the vibrational mode structure of our chains. We numerically calculate the normal mode frequencies and eigenvectors for each unique ordering of the ions in the chain. To find the radial normal modes frequencies we first consider only the axial direction ($z$-axis). We minimize the potential energy of the ions to find the equilibrium ion positions in axial direction assuming $x$ and $y$ position of the chain to be zero. One might worry about incorrectly restricting the chain to one dimension, but we can observe and avoid the zig-zag transition in our calculation because it is marked by the softening of the lowest transverse vibrational mode to zero frequency. Once we find the equilibrium positions, we expand the full trap potential around these equilibrium ion positions to find the normal mode frequencies and eigenvector magnitudes. \linebreak    
    \indent An example of the numerical calculation for the eigenvector magnitudes as a function of the trap aspect ratio in a chain of two barium and two ytterbium ions is shown in Figure \ref{fig:normal_modes}. Here the chain order is Ba-Yb-Ba-Yb (BYBY), and we plot the eigenvector magnitudes for the second barium ion $x$-direction (see Figure \ref{fig:normal_modes},a). The strong dependence of the normal mode coupling on the trap aspect ratio can be seen. The eigenvector magnitudes approach 0.5 (which, for a four-ion chain, corresponds to the maximum coupling) when aspect ratio decreases. The same trend is seen for ytterbium ions. The table in Figure \ref{fig:normal_modes},b summarizes the eigenvector magnitudes of the first ytterbium ion and the second barium ion in the BYBY ordering for the trap aspect ratios of 2.9 and 5.5. These numerical results lead to the important conclusion: better coupling can be achieved by lowering trap aspect ratio. Thus, measuring radial mode occupation numbers at different trap aspect ratios is of interest.

        %%%%Ba levels%%%%

%%%%%%%EIGENVECTOR SCALING WITH FREQ%%%%%%%
    
    \section{Apparatus and experimental procedure}

    The mixed species chain measurements are carried out in a linear RF trap similar to that described in~\cite{PhysRevA.81.052328}. Species-selective photoionization techniques are used to load both $^{138}$Ba$^{+}$ and $^{174}$Yb$^{+}$, where barium is Doppler cooled on the $6S_{1/2}-6P_{1/2}$ transition with a 986~nm Extended Cavity Diode Laser (ECDL) frequency doubled to produce 493~nm light.  The $6P_{1/2}$ level decays to the metastable $5D_{3/2}$ level with a branching ratio of 0.25, so a 650~nm ECDL is used to pump the population out of this state. An applied 5~gauss magnetic field prevents the creation of dark states. All relevant transitions and energy levels of Ba$^+$ are shown in Figure \ref{fig:BaLevels}. Ytterbium is not directly laser-cooled, but sympathetically cooled by Coulomb interactions with the cold barium ions in the chain. 

    All temperature measurements are performed on barium ions using a 1762~nm fiber laser (Koheras Adjustik) stabilized to a Zerodur-spaced reference optical cavity with a free spectral range of 500~MHz and finesse of 1000. A complete description of the apparatus can be found in~\cite{PhysRevA.81.052328,Wright:2016:TSQ:3022910.3022945}. 

    While the short-term linewidth of the 1762~nm laser is of order 100~Hz~\cite{PhysRevA.85.023401}, slow frequency drifts result in a 5~kHz linewidth. This is consistent with the locking system being incapable of stabilizing the laser frequency to better than a few kHz, while the laser itself has a narrow linewidth. Our frequency scans take tens of minutes to a few hours and are thus broadened, so features separated by $<$15~kHz are not well resolved.
    
    The 1762~nm laser is aligned perpendicular to the trap $z$-axis, and at roughly 45 degrees to the $x$- and $y$-axes. Thus, only the radial sidebands are present in the frequency scans. The laser is focused to a 30~$\mu$m Gaussian spot size centered at the ion chain, driving all ions with comparable Rabi frequencies. Polarization control of the 493~nm cooling laser beam with a Pockel's cell is used to optically pump barium ions to the same $6S_{1/2}$ Zeeman state at the start of each run. After the 1762~nm laser exposure, we detect the state of each barium ion simultaneously with an Electron-Multiplied Charge-Coupled Device (EMCCD) camera (Andor Luca) by imaging light scattered by the ions on the cooling transition. If the $6S_{1/2}-5D_{5/2}$ transition took place, then the ion appears dark; otherwise the ion appears bright. At the end of each experimental run, a short pulse of 614~nm laser light from a frequency-doubled ECDL near 1228~nm returns the barium ions to the ground state. 

    The experimental sequence is as follows. We Doppler cool the barium ions with the 493~nm and 650~nm lasers for approximately 50~ms with the Pockel's cell turned to a high voltage. Then we discharge the Pockel's cell to curcularly polarize the 493 nm laser. The switching time of the Pockel's cell is approximately 1~ms, so we wait for 5~ms to ensure full optical pumping. The 493~nm and 650~nm laser beams are then extinguished successively, and a 1762~nm pulse is applied. Afterward, the 493~nm, 650~nm, and Pockel's cell are turned on for 50~ms to read out the states of the barium ions. The experiment is repeated, varying either the frequency of the 1762~nm during sideband scans, or the duration of the pulse during Rabi flop experiments.
    
    During the experiment the chain sometimes spontaneously reorders, due to either background gas collisions or high chain temperature. When we detect reordering, we discard the experimental cycle and re-establish the chain order by performing melting and crystallization cycles induced by shuttering and unshuttering the Doppler cooling lasers.

     \section{Results}
  
    \begin{figure*}[p!]
        \includegraphics[width=1\linewidth]{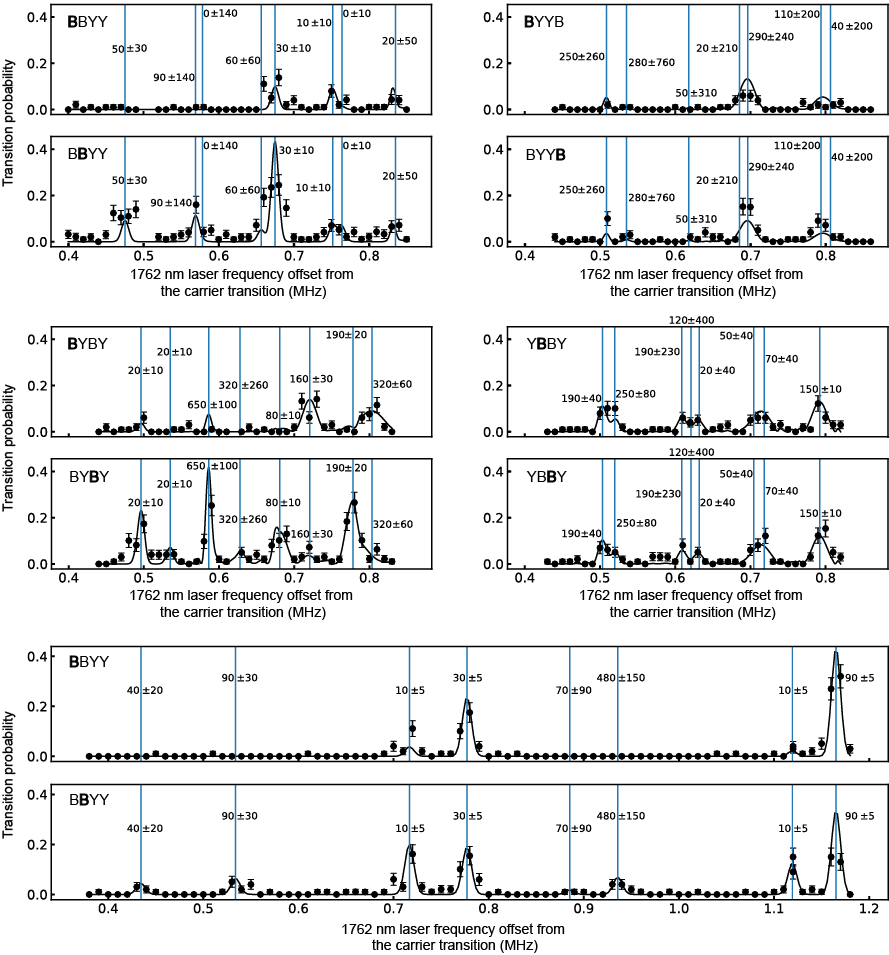}
        \caption{(Color online) 1762~nm laser scans showing $\Delta n = 1$ sideband transitions. Two sets of radial mode peaks are present in each scan. Bold font is used to indicate which Ba$^+$ in the chain the spectrum was taken from. Radial mode occupation numbers are extracted from the peak amplitudes. The vertical lines indicate the calculated frequencies of the normal modes, and the associated radial mode occupation numbers are printed close by. The error bars on the measured occupation numbers are statistical. The Doppler limit for the secular frequency modes between 0.5-0.9~MHz corresponds to an $\bar{n}$ of $16-30$ for a $^{138}$Ba$^+$. The peak sizes are proportional to the associated $\bar{n}$'s, the eigenvector component $\beta_{ij}$'s, and the exposure time, which is not displayed but is adjusted to stay in the weak excitation regime. In the bottom scan, a squeeze potential is applied to the trap's RF rods, which pushes the radial mode frequencies apart and changes the eigenvector component sizes along the two axes. 
         }
         \label{sideband_scans}
    \end{figure*}
    
            \begin{figure}
        \includegraphics[width=1\linewidth]{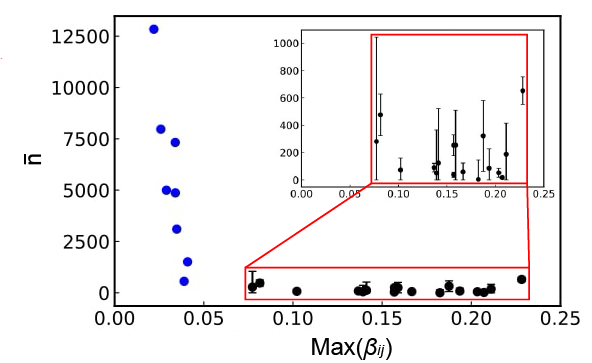}
        \caption{(Color online)
	    The $\bar{n}$ values vs maximum eigenvector component. The normal mode occupation number shows threshold-like behavior. 
	    All new data in the eigenvector component range $\beta_{ij}>0.05$, shown in black, were taken at an aspect ratio of 2.9 except for the last set of data in Figure \ref{sideband_scans} corresponding to the Ba-Ba-Yb-Yb configuration. Data in the range $\beta_{ij}<0.05$, shown in blue, are taken from~\cite{UW} and are not presented in Figure \ref{sideband_scans}. These data do not have uncertainty displayed because the measured temperatures are outside the Lamb-Dicke regime.
         }
         \label{nbar_vs_eig_both}
    \end{figure}

	We perform weak excitation sideband scans over all radial modes of all possible configurations of two barium and two ytterbium ions. The results are summarized in Figure~\ref{sideband_scans}. The normal mode decompositions are used to generate the data fits, as the peak frequencies are the frequencies of the radial eigenmodes, and the relative peak amplitudes on different ions are given by the eigenvector components and the average occupation numbers (see eqn. \ref{eqn:FullWeakExcite}). The 1762~nm carrier Rabi frequency at each ion position is measured in a separate experiment by performing the carrier Rabi flops with a chain of four barium ions. 
    
    The data from our earlier experiments (which can be found in~\cite{UW} and are not shown in Figure~\ref{sideband_scans}) indicated the $\bar{n}$ to be 50 to 2000 times higher than the Doppler limited $\bar{n}$ for some of the normal modes, which meant that chains under those conditions could not be used successfully for sympathetic cooling or inter-species quantum logic gates. Analysis of that data revealed a correlation~\cite{UW} between the maximum eigenvector component of the cooled ion species and the measured $\bar{n}$. Noting the correlation, we searched for trap and chain configurations for which the eigenvector components would be large for the cooled species in all modes. Using the numerical normal mode structure calculation tool described in the theory section, we found that lowering the trap aspect ratio from 5.5 (used in \cite{UW}) to 2.9 increases the normal mode eigenvector component values above the expected 0.05 \textquotedblleft threshold~\textquotedblright value regardless of chain ordering. Some chain orderings were found to be better cooled than the others. For example, in the BYYB configuration, the maximum eigenvector component is at least 30\% lower than the maximum eigenvector component in any other configuration. 

	We summarize our measurements of the radial mode $\bar{n}$ as a function of the maximum eigenvector component for both the old set~\cite{UW} with trap aspect ratio of 5.5 and the new set of data with trap aspect ratio of 2.9 in Figure~\ref{nbar_vs_eig_both}. We find very large reduction in $\bar{n}$ for maximum eigenvector component above approximately 0.05. These values of $\bar{n}$ correspond to normal mode temperatures between the Doppler limit for a single trapped barium ion and about 40 times that Doppler limit. These values are consistent with the technical difficulties of achieving the Doppler cooling limit in a lambda system such as Ba$^+$~\cite{Janik:85}. At low intensities of the cooling and repump lasers, the rate of heating from residual micromotion balances the cooling rate well above the Doppler limit, while at high intensities the combination of power broadening and interference effects between the cooling and the repump transitions causes the minimum temperature to also increase well above the Doppler limit.\linebreak

    \section{Conclusions}
    We studied the sympathetic cooling of barium-ytterbium chains in a linear RF trap in which only barium is directly laser cooled. We found that reducing the aspect ratio of the trap reduced the measured ion temperature in all normal modes. The ion chain has been cooled from a factor of 50 to 2000 the Doppler limit (well outside the Lamb-Dicke regime) to within a factor of 1 to 40 the Doppler limit, which is consistent with the cooling limitations in our setup. Our measurements show that the radial normal mode temperature is strongly dependent on the size of the maximum normal mode component of the cooled species.
    We conclude that in order to achieve efficient sympathetic cooling and interspecies quantum logic gates using radial vibrational modes, traps with lower aspect ratios are desirable. \linebreak 
    
    \begin{acknowledgments}
The authors wish to thank John Wright and Wen-Lin Tan for help with earlier parts of the experiments, and Megan Ivory, Jennifer Lilieholm, Alexander Pierce, Ramya Bhaskar and James Walker Steere for helpful discussions. This research was supported by National Science Foundation Grant No. 1505326.\linebreak
    \indent Tomasz P. Sakrejda and Liudmila A. Zhukas contributed equally to this manuscript.
    \end{acknowledgments}

    \bibliography{cut_down_version}

\end{document}